\documentclass[journal,10pt,twocolumn,letterpaper]{IEEEtran}

\usepackage{amsmath,amssymb}
\usepackage{graphicx}
\usepackage{bm}
\usepackage{hyperref}
\usepackage[utf8]{inputenc}
\usepackage{algorithm}
\usepackage{algpseudocode}
\usepackage{booktabs}
\usepackage{multirow}
\usepackage{etoolbox}
\usepackage{subfig}

\AtBeginEnvironment{tabular}{\footnotesize}
\begin{document}

\title{Efficient Audio Enhancement with a Differentiable Psychoacoustic Loss}

\author{Wallace~Abreu,
        Bernardo~V.~Miranda,
        and~Luiz~W.~P.~Biscainho%
\thanks{W. Abreu, B. V. Miranda, and L. W. P. Biscainho are with SMT, DEL/Poli \& PEE/COPPE, Federal University of Rio de Janeiro, Brazil (corresponding e-mail: wallace.abreu@smt.ufrj.br).}%
\thanks{B. V. Miranda is also with LCTI, Télécom Paris, IP Paris, France.}%
\thanks{Author's Accepted Manuscript (AAM). Published in the Journal of the Audio Engineering Society (JAES). DOI: 10.17743/jaes.2022.0277}
}

\markboth{Author's Accepted Manuscript --- Published in Journal of the Audio Engineering Society}{}

\maketitle

\begin{abstract}
Audio enhancement consists of improving the perceived quality of audio signals. Initially, with the aim of addressing bandwidth extension, this work proposes $\textrm{AEROMamba}_{\textrm{P}}$, an efficient variant of the AERO super-resolution architecture where attention and LSTM layers are replaced by the Mamba state-space model, and which incorporates a newly developed differentiable perceptual loss derived from the Perceptual Audio Quality Measure (PAQM). During training, the architecture requires approximately 2–4x less GPU memory than the baseline; during inference, it achieves a 14x speedup while using only one-fifth of the GPU memory. When upsampling both a piano dataset and MUSDB18 from 11.025 kHz to 44.1 kHz, subjective listening tests show that $\textrm{AEROMamba}_{\textrm{P}}$ outperforms AERO by 15\% in perceived quality scores. Next, to handle the enhancement of audio signals that have been highly compressed by lossy coding, it is further proposed $\textrm{AEROMamba}_{\textrm{P} \bar{\textrm{S}}}$, which applies the same framework but replaces STFT reconstruction losses with the PAQM loss, specifically to enhance MP3 encoded audio at 32 kbps. In listening evaluations, $\textrm{AEROMamba}_{\textrm{P} \bar{\textrm{S}}}$ achieves 52\% higher quality rating than $\textrm{AEROMamba}_{\textrm{P}}$ when restoring compressed audio. These results demonstrate that PAQM-driven training coupled with lightweight state-space modeling yields high perceptual quality and computational efficiency in both band-limited and compressed audio scenarios.
\end{abstract}

\maketitle
\section{INTRODUCTION}
Since the development of sound recording devices in the 19th century~\cite{IASA14}, audio signals have been dealt with in a multitude of applications, especially in communications and entertainment sectors. Technology in these fields has evolved to meet the main requirements of each specific application, which can be different due to their own concept of quality. For example, in speech systems, such as telephony, providing intelligibility is the main goal, whereas the target in general audio devices is fidelity~\cite{Coelho17}. 

The frequency span of human hearing extends approximately from 20 Hz to 20 kHz~\cite{Bosi02} (for tones), requiring that high-fidelity systems of audio recording, reproduction, and their storage media operate at least in that frequency range. However, especially for analog audio, factors such as limitations in technology or degradation of physical media can affect the preservation of full-range spectrum~\cite{Copeland08}. 

In digital audio, transmission and storage costs can be reduced through data compression, for which the simplest solution is direct sample discarding (decimation). This procedure requires prior low-pass filtering to prevent aliasing, directly reducing the maximum frequency present in the resulting signal. If a high downsampling factor is used, even midrange frequencies (the most important ones) can be cut. Another common method to reduce storage requirements for digital audio is to use lossy perceptual audio coding~\cite{Blauert1997} algorithms, such as MP3~\cite{Brandenburg99}. In this case, the overall frequency content is quantized according to a psychoacoustic model for minimum perceptual degradation. This procedure has a more significant impact at high frequencies, even limiting bandwidth if the target bitrate requires it, but low frequencies can also be considerably affected.

Solutions based on classical signal processing that address these degradations within certain limits have been developed and made available in the literature. An extensive study of audio (including music and speech) bandwidth extension (BWE) targeting different applications can be found in~\cite{Larsen04}, encompassing nonlinear transformations and linear filtering. In particular, spectral band replication (SBR)~\cite{Ekstrand02} has been shown to enable compressed signal reconstruction with extended bandwidth at decoding. A comprehensive study on BWE in the context of speech telephony is provided in~\cite{Schmidt08}, emphasizing solutions using source-filter modeling. In this context,~\cite{Spanias06} explores psychoacoustic criteria for improved naturalness. However, in the case of audio encoded (lossily) at very low bit rates, undoing the nonlinear distortions that have been imposed on the signal content according to perceptual criteria involves a coherent reconstruction of the entire spectrum. And the ``intelligent'' decisions needed to approach this task exceed the capabilities of traditional signal processing.

As a note, the term \emph{super-resolution} is often related to audio BWE~\cite{Mandel22, Chen24, Choi22}, and this work follows the same terminology. In turn, the process of inverting lossy coding is referred to here as \emph{ encoded audio enhancement}.

Techniques based on neural networks (NNs) became the state of the art in audio super-resolution, operating on raw  waveforms~\cite{Kuleshov17} or in the spectral domain~\cite{Mandel22, Valimaki22, Lagrange20}. Initially, these models used feedforward architectures with standard supervised learning strategies, but subsequently evolved into solutions built as a combination of modules, such as convolutional and recurrent networks, and following different paradigms, such as generative adversarial networks (GANs)~\cite{Mandel22,Li2025} or diffusion models~\cite{Chen24}.  

In~\cite{Mandel22}, the authors introduce AERO, an encoder-decoder NN implemented within a GAN framework, designed for 4-fold upsampling of music and speech signals having particular initial sampling frequencies. AERO showcases the importance of adversarial training for natural-sounding BWE and achieves higher scores than later methods such as the zero-shot diffusion-based network BABE~\cite{Moliner24} in BWE tasks. However, AERO is computationally demanding, both in training and inference, mainly because of its use of recurrent and attention layers. These issues are alleviated in AEROMamba~\cite{Abreu2024lamir}, where the mentioned layers are replaced by the state-space model Mamba~\cite{Gu2024mamba}, yielding better performance without degrading the perceptual quality in music enhancement.

Another noteworthy deep learning model for BWE is AudioSR~\cite{Chen24}, which, unlike AERO, is capable of enhancing signals with any initial sampling frequency up to the sampling rate of 48kHz. The authors demonstrate the capability of diffusion models to enhance audio signals of different natures, such as music, speech, and sound effects, to a very high quality. As a diffusion model, AudioSR has a slow inference procedure as a significant drawback, and is unsuitable for real-time applications. 

Finally, the (recent) Apollo~\cite{Li2025} model is also an important contribution in the field of audio enhancement. This model employs RoFormers~\cite{Su2024} and Temporal Convolutional Networks (TCNs)~\cite{Lea2017} to generate features in time for different frequency sub-bands. These features are used to restore signals compressed by MP3 at 32 and 48 kbps. As AERO, Apollo includes attention modules that increase its computational cost considerably. This is illustrated by the use of eight NVIDIA RTX 4090 GPUs for its training. 

AERO, AudioSR, and Apollo are included here as state-of-the-art models for music enhancement. However, their performances are not compared (which, incidentally, is also not done in their respective publications).

This paper introduces $\textrm{AEROMamba}_{\textrm{P}}$, a first-of-its-kind audio super-resolution network, to the best of the author's knowledge, that includes a model-based perceptual quality evaluation in its loss function, namely the Perceptual Audio Quality Measure (PAQM)~\cite{Beerends1992}. This novelty is enabled by the development of a differentiable implementation of PAQM that is appropriate as a training loss for deep learning models.

The proposed model is evaluated using objective measures as well as subjective listening tests in two different applications. The first experiment concerns super-resolution of bandlimited audio: the performance of $\textrm{AEROMamba}_{\textrm{P}}$ is compared to AERO and AEROMamba when upsampling test signals from 11.025 kHz to 44.1 kHz, also including a pre-trained AudioSR model in the objective evaluation. The second experiment deals with the enhancement of highly compressed audio, comparing the effectiveness of AEROMamba, $\textrm{AEROMamba}_{\textrm{P}}$, and $\textrm{AEROMamba}_{\textrm{P} \bar{\textrm{S}}}$ (a variation of the latter) in undoing the losses caused by MP3 audio encoding at 32 kbps (which also reduces the bandwidth to 11.025 kHz). The tests were based on music signals from two databases: PianoEval (homemade) and MUSDB. To ensure reproducibility and allow perceptual examination, a public repository and a webpage\footnote{aeromamba-paqm.github.io} accompany this work, containing the source code, model checkpoints, and audio examples, as well as supplementary materials for the conducted experiments.

The paper is organized as follows. Section 1 presents the AERO NN architecture. Section 2 introduces deep state space models, focusing on the Mamba architecture. Section 3 outlines the PAQM. Section 4 details the proposed models (AEROMamba and its variants). Section 5 describes the datasets used in the experiments. Section 6 discusses the evaluation methodologies. Sections 7 and 8 present experimental setup and results for super-resolution of bandwidth-reduced and enhancement of highly compressed audio, respectively. Finally, Section 9 concludes the paper.

\section{AERO}
\subsection{Architecture}
AERO~\cite{Mandel22} is an audio super-resolution model inspired by Demucs~\cite{Defossez22}, a NN architecture for music source separation. Given a signal $\bm{y} \in \mathbb{R}^T$ and its downsampled version $\bm{x}_l \in \mathbb{R}^{T/s}$, where $T$ is the original signal length and $s$ is a fixed integer downsampling factor, AERO aims to generate $\hat{\bm{y}} \approx \bm{y}$, reconstructing the lost high-frequency content of $\bm{x}_l$. For speech signals, Mandel et al.~\cite{Mandel22} trained / tested the model on VCTK~\cite{Veaux17} in various settings. For music, they used MUSDB~\cite{musdb18-hq} in the $11.025 \to 44.1$ kHz setting.

\begin{figure}[!ht]
    \centering
    \includegraphics[width=76mm]{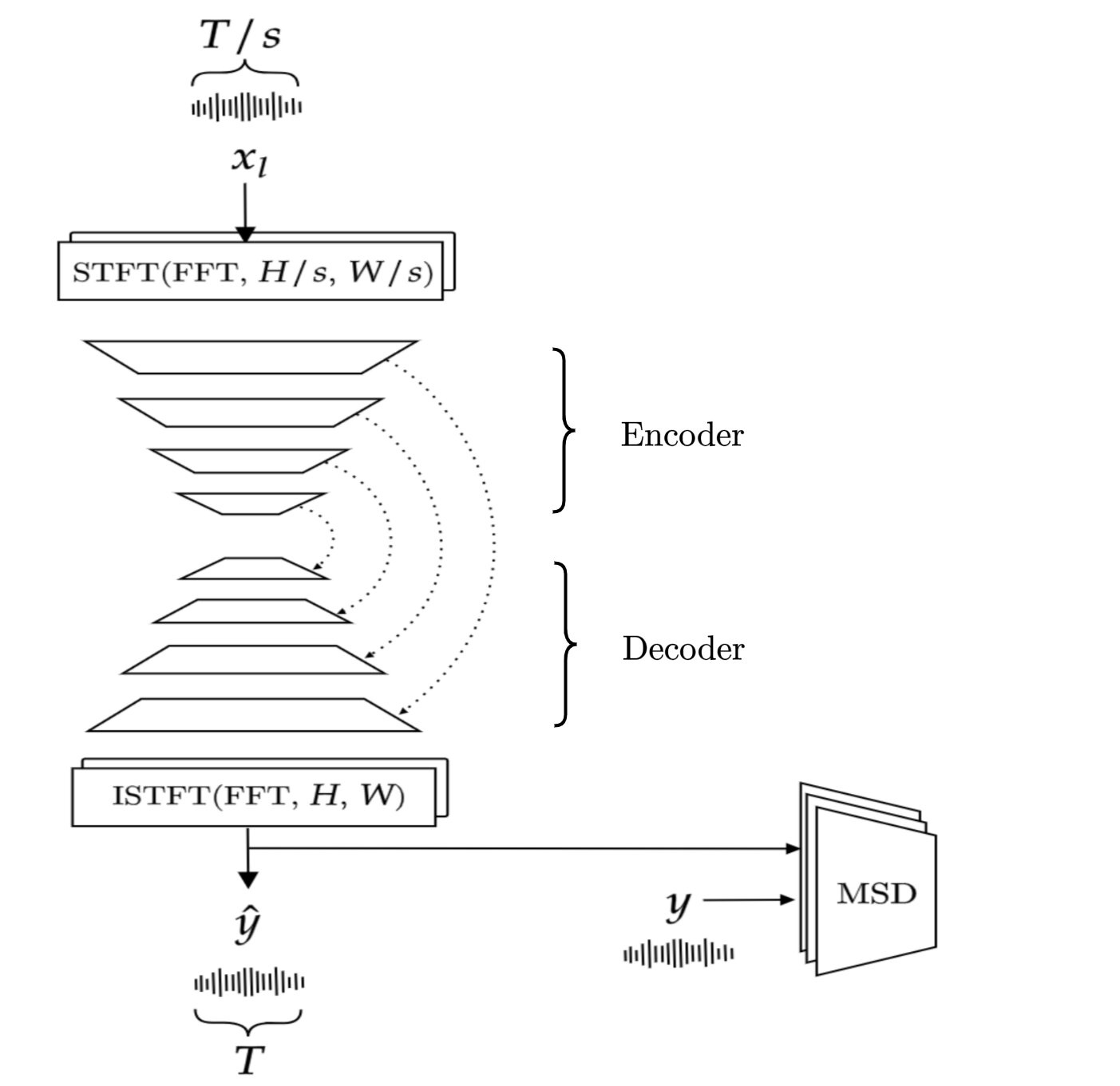}
    \caption{AERO architecture.}
\label{fig:aero}
\end{figure}

As shown in Figure~\ref{fig:aero}, the AERO architecture is composed of a generator based on a U-Net and a multi-scale discriminator. In the generator (left-hand side), the input audio $\bm{x}_l$, with $T/s$ samples, is transformed to a complex spectrogram $\bm{X}_l$ with $F$ frequency bins using a window size $W = F/s$ and hop length $H/s$, where $s$ is a downsampling factor. The window is zero-padded to ensure $F$ bins as $W < F$. The NN processes the low-resolution spectrogram $\bm{X}_l$, yielding a high-resolution spectrogram $\hat{\bm{Y}}$ to match $\bm{X}_l$, filling high-frequency content. The network treats the real and imaginary components of the spectrogram as separate channels, thus jointly estimating the magnitude and phase. An inverse STFT with $F$ FFT points, window size $sW$, and hop size $H$ then reconstructs the signal $\hat{\bm{y}}$. During training, the generated and target signals are fed into multi-scale discriminators (MSD)~\cite{Kumar19} trained with time-domain adversarial losses, feature losses, and spectral reconstruction losses. 
\begin{figure}[!ht]
    \centering
    \includegraphics[width=65mm]{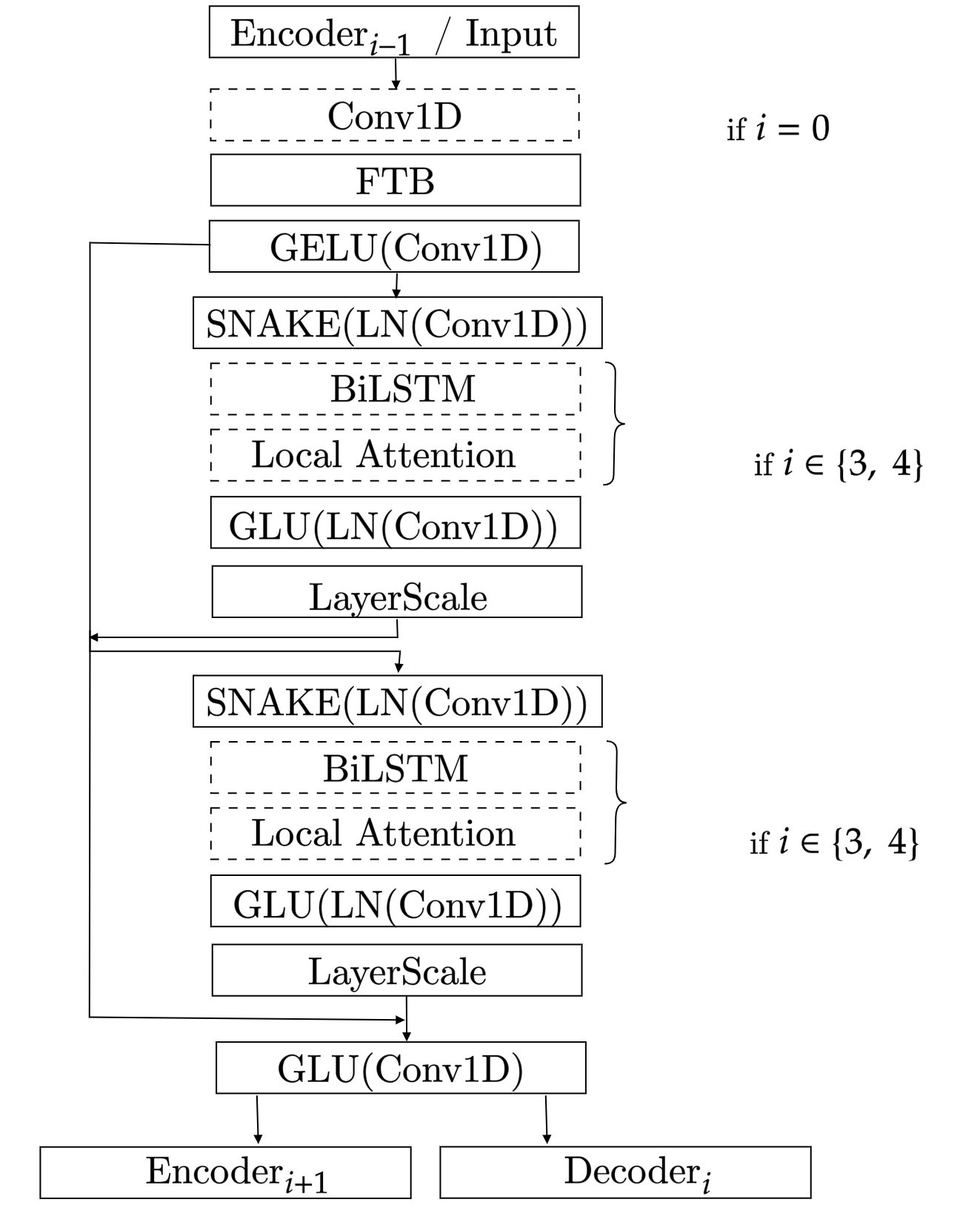}
    \caption{AERO encoder layer.}
\label{fig:aero_enc}
\end{figure}
Fundamentally, AERO maps a zero-padded spectrogram to a high-resolution spectrogram via a NN transformation that is similar to interpolation in the frequency axis, with both the input and output having the same shapes. This procedure contrasts to other common methods, such as resampling prior to a spectral content fill~\cite{Li21}, or generating a separated high-frequency spectrum that is concatenated with the low-frequency spectrogram~\cite{Hu20} --- a strategy that often creates artifacts at the frontier of the union.

The AERO U-Net encoder is shown in Figure~\ref{fig:aero_enc}. In the first stages of the encoder, the complex-by-channels spectrum of the input signal is modified through 1D convolutions in frequency followed by a Frequency Transformer Block (FTB)~\cite{Yin19}, essentially an attention layer that helps detect non-local correlations in the frequency axis. This allows the network to have a larger receptive field over the frequency axis thanks to the attention mechanism (in a fully convolutional architecture, the receptive field from relatively small kernels should cover the whole spectrum). Its output is processed by a 1D convolutional layer followed by a GELU~\cite{Hendrycks2023} non-linearity.

The core of the encoder is made of two identical residual blocks composed of layer normalization (LN), Conv1D, BiLSTM (Bidirectional Long Short-Term Memory layer), local attention and GLU~\cite{Dauphin17}. They are preceded by 1D convolutional layers followed by LN and a Snake~\cite{Ziyin20} activation function. LN reduces training time through recentering and rescaling of the input data~\cite{Ba16}, and Snake has been shown to be effective for periodic data~\cite{Ziyin20}. A distinction between AERO and Demucs lies in their choice of activation functions: AERO employs GLU and Snake, while Demucs uses only the first. A layer scale (LS)~\cite{Touvron21} operation follows each residual block to enhance model accuracy and convergence by rescaling residual layer outputs with learnable parameters.

Intuitively, the encoder architecture tries to capture local spectral patterns through 1D convolutions. However, as frequency components can be influenced by bins outside of their close neighborhood, FTB and BiLSTMs are essential to include long-range memory and identify spaced features in the time and frequency dimensions. 

The AERO decoder is symmetrical to the encoder and shares skip connections with the encoder layers. AERO employs a MelGAN~\cite{Kumar19} multi-scale discriminator for adversarial training: an ensemble of three independent discriminators, each of them a set of strided convolutional layers operating directly on the waveform. The strided convolutions act as downsamplers and allow the discriminators to focus on patterns in different spectral regions.

\subsection{Training}
The training objective for the discriminator is the MelGAN~\cite{Kumar19} hinge loss
\begin{align}
    L_{\mathcal{D}} &= \mathbb{E}_{\bm{y}} \left[ \frac{1}{R} \sum_{r}  \max \left(0, 1 - \mathcal{D}_{r}(\bm{y})\right) \right] \\
                    &+ \mathbb{E}_{\hat{\bm{y}}} \left[ \frac{1}{R} \sum_{r}  \max \left(0, 1 + \mathcal{D}_{r}(\hat{\bm{y}})\right) \right],
\end{align}
where $r$ indexes each of the $R$ discriminator resolutions. For the generator, the training cost function has three components: a multi-resolution STFT reconstruction loss $L_\text{rec}$, an adversarial loss $L_\text{adv}$, and a feature map loss $L_{\text{fmap}}$. 

The multi-resolution reconstruction loss 
\begin{equation}
    L_{\mathrm{rec}}=\mathbb{E}_{\bm{y}, \hat{\bm{y}}}\left[\frac{1}{N_w} \sum_{w=1}^{N_w}\left(L_{\mathrm{sc}}^{(w)}+L_{\mathrm{mag}}^{(w)}\right)\right]
\end{equation}
is the expected value, considering true and generated samples, of the average sum of the spectral convergence loss
\begin{equation}
    L_{\mathrm{sc}}^{(w)}=\frac{\||\bm{Y}^{(w)}|-|\hat{\bm{Y}}^{(w)}|\|_{\mathrm{F}}}{\||\bm{Y}^{(w)}|\|_{\mathrm{F}}},
\end{equation}
where $||\cdot||_F$ is the Frobenius norm, and the log magnitude loss
\begin{equation}
    L_{\mathrm{mag}}^{(w)}=\frac{1}{N_w}\|\log|\bm{Y}^{(w)}|-\log|\hat{\bm{Y}}^{(w)}|\|,
\end{equation}
both computed for a specific window size $w$, over all $N_w$ window sizes $w$. The adversarial generator loss 
\begin{equation}
    L_{\text{adv}} = \mathbb{E}_{\hat{\bm{y}}} \left[ \frac{1}{R} \sum_{r}  \max \left(0, 1 - \mathcal{D}_{r}(\hat{\bm{y}})\right) \right]
\end{equation}
is a standard GAN generator loss. And the feature map loss
\begin{equation}
    L_{\text{fmap}} = \mathbb{E}_{\hat{\bm{y}}} \left[ \frac{1}{RL} \sum_{r,l} \frac{1}{T_{r,l}}\sum_{t} \left|\mathcal{D}_{r, t}^{(l)}(\bm{y}) - \mathcal{D}_{r, t}^{(l)}(\hat{\bm{y}}) \right| \right],
\end{equation}
where $L$ is the number of layers of the discriminators, $\mathcal{D}_{r, t}^{(l)}$ is the $t$-th output of the discriminator $r$ at layer $l$, and $T_{r,l}$ denotes the length of the layer in the time dimension, is inspired by~\cite{Li21}. It compares the feature maps of the layers of the discriminators for true and generated samples. 

The expression for the total generator loss is
\begin{equation}\label{eq:aero_g_loss}
    L_{\mathcal{G}} = L_\text{adv} + L_\text{rec} + \lambda L_{\text{fmap}},
\end{equation}
where $\lambda$ is a scaling factor. Note that the reconstruction cost operates only on the magnitude spectrogram, but the adversarial and feature costs also penalize phase mismatches, as they act on the true and estimated waveforms.

\section{MAMBA}
\subsection{Motivation}
Despite the remarkable success of Transformers~\cite{Vaswani17} in modeling sequential data, their self-attention mechanism suffers from an intrinsic limitation: the context length is finite and often fixed by design. This restricts the ability of the model to capture dependencies across arbitrarily long sequences since the memory of the model is bounded by the size of the attention window. Additionally, the quadratic computational and memory cost of the attention mechanism with respect to the sequence length makes Transformers inefficient for long or continuous signals~\cite{Gu2024mamba}, such as audio or time series data.

In contrast, recurrent neural networks (RNNs)~\cite{Goodfellow14} naturally process sequences of arbitrary length, maintaining a hidden state that theoretically allows for infinite context. Their recursive formulation provides a memory-efficient way to represent temporal dependencies, making them appealing for sequential data modeling. However, standard RNNs are limited in preserving long-range dependencies (LRD) over extended time horizons, as gradients tend to vanish or explode through time~\cite{Pascanu13}. Moreover, traditional RNNs lack selectivity, missing input-dependent mechanisms to most effectively use past information. 

In response to the mentioned limitations, various deep state space models (SSMs) have been developed to address each of the specified challenges progressively. These models include the Linear State-Space layer (LSSL)~\cite{Gu21LSSL}, Structured State-Spaces (S4)~\cite{Gu22S4}, and Mamba~\cite{Gu2024mamba}.

\subsection{Linear state-space layer}
Linear State-Space layers (LSSLs)~\cite{Gu21LSSL} present a compromise between convolutional models, which are efficient yet constrained in context, and RNNs, which provide infinite context but exhibit instability. LSSLs are derived from the state-space representation of linear dynamical systems as seen in control theory. In continuous time, the operation of a linear time-invariant system can be described as
\begin{align}\label{eq:ct-state-space}
        \dot{\bm{q}}(t) = \bm{A} \bm{q}(t) + \bm{B}\bm{x}(t), \\
        \bm{y}(t) = \bm{C} \bm{q}(t) + \bm{D}\bm{x}(t),
\end{align}
where $\bm{q}(t)$, $\bm{x}(t)$ and $\bm{y}(t)$ are respectively the state, input and output vectors, containing the processed signals; and $\bm{A}$, $\bm{B}$, $\bm{C}$ and $\bm{D}$ are respectively the state, input, output and feed-forward matrices that describe the system. 

The LSSL uses the discrete-time version of this formalism in a deep learning framework: the matrices $\bm{A}$, $\bm{B}$ and $\bm{C}$ become learnable parameters optimized via gradient descent. This provides a differentiable linear dynamical system that models temporal dependencies efficiently. However, since the solution to the system is a function of exponentials of $\bm{A}$, LSSLs still suffer from exponential decay or growth, leading to vanishing or exploding gradients in long sequences~\cite{Gu22S4}. Therefore, while LSSLs improve stability and interpretability, they still ``forget'' distant information.

\subsection{Structured State-Spaces (S4)}
To mitigate the forgetting problem, S4~\cite{Gu22S4} introduces a way to choose the state transition matrix $\bm{A}$ based on high-order polynomial projection operators (HiPPOs). These operators impose a structure on matrix $\bm{A}$ that improves the model’s ability to retain past information, effectively extending its memory horizon. Intuitively, HiPPOs maintain a compressed summary of all past inputs, enabling the system to approximate functions of the entire input history.

Since the theoretical background of S4 is set over continuous systems, its implementation requires discretization with some step $\Delta$ (e.g. via bilinear transform) into the form 
\begin{align}\label{eq:bilinear-state-space}
        \bm{q}[n] &= \overline{\bm{A}} \bm{q}[n-1] + \overline{\bm{B}}\bm{x}[n], \\
        \bm{y}[n] &= \overline{\bm{C}} \bm{q}[n] ,
\end{align}
with 
\begin{align}\label{eq:discretization}
    \overline{\bm{A}} &= (\bm{I} - \Delta/2 \cdot \bm{A})^{-1} (\bm{I} + \Delta/2 \cdot \bm{A}), \\
    \overline{\bm{B}} &=  (\bm{I} - \Delta/2 \cdot \bm{A})^{-1}\Delta \bm{B}, \\
    \overline{\bm{C}} &= \bm{C}.
\end{align}

S4 further reformulates the recurrence of~\eqref{eq:bilinear-state-space} as a time convolution, enabling parallelization and efficient computation through the Fast Fourier Transform (FFT). This would normally require calculating specific convolution kernels, increasing the computational overhead of using these state-spaces. However, a key innovation of S4 is the use of a normal plus low-rank (NPLR) decomposition of $\overline{\bm{A}}$, which alleviates this burden.

\subsection{Selective State-Spaces (Mamba)}
While S4 captures long-range dependencies efficiently, it remains a linear time-invariant model. This can be a fundamental limitation in modeling data that comes from nonlinear and time-variant phenomena, where the importance of past information changes dynamically with the context.

In contrast to S4, Mamba~\cite{Gu2024mamba} introduces a parameterization of $(\Delta, \bm{B}, \bm{C})$ with respect to $\bm{x}$.
Although $\bm{A}$ is not explicitly parameterized, it interacts with the input through the parameter $\Delta$ of the discretization procedure.

This mechanism introduces selectivity: the model learns to focus on relevant inputs while ignoring irrelevant ones, effectively emulating the context-dependent weighting of attention mechanisms. Unlike attention, however, Mamba maintains constant-time inference and linear-time training, owing to a hardware-aware implementation that performs all recurrence computations within the GPU’s fast memory (SRAM) using a parallel scan algorithm~\cite{Gu2024mamba}. This combination of efficiency and selectivity allows Mamba to achieve attention-like modeling capacity without the related computational burden.

\begin{figure}[H]
    \centering
    \includegraphics[width=60mm]{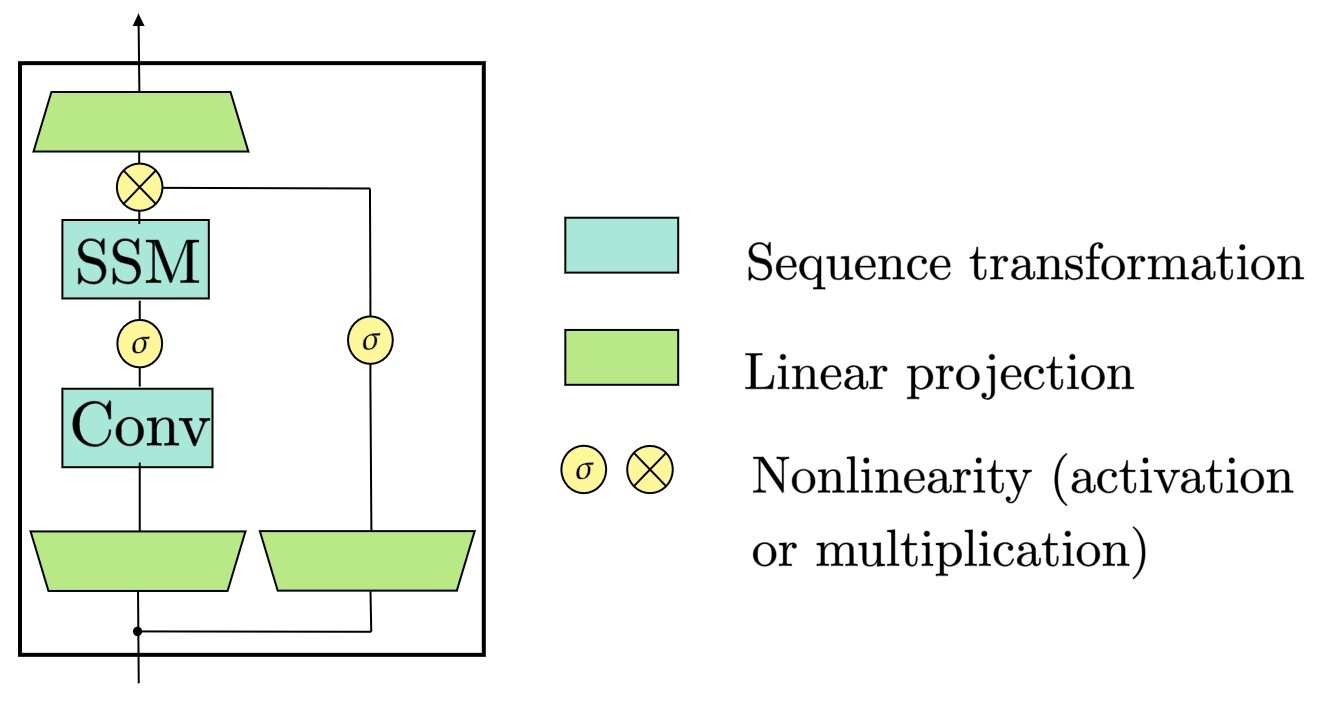}
    \caption{Mamba block architecture.}
    \label{fig:mamba}
\end{figure}

The final product of Selective State-Spaces is the Mamba architecture, shown in Figure~\ref{fig:mamba}: a combination of linear projections, a convolutional layer, a residual connection, a selective SSM, and SiLU/Swish~\cite{Ramachandran17} functions as nonlinear activations. 

\section{PERCEPTUAL AUDIO QUALITY MEASURE (PAQM)}\label{sec:PAQM}

Audio quality measures based on models of human auditory perception correlate better with subjective quality assessment than simple signal-based measures such as signal-to-noise ratio (SNR), and are the standard for audio quality evaluation since the 1990s~\cite{Torcoli21}.

The perceptual audio quality measure (PAQM) is one of the first perceptually inspired objective audio quality measures. Introduced in~\cite{Beerends1992} in the context of the evaluation of digital audio codecs, it is based on calculating the difference between the internal --- or psychoacoustic --- representations of a reference and a test signal, as follows. 

First, a time-frequency representation of the signal is obtained using a short-time Fourier transform (STFT). A pitch-based representation is then obtained by mapping the frequency axis of the STFT from hertz to barks. The bark scale models human perception of pitch and is the psychoacoustic equivalent of the frequency scale. 

After this, the bark spectrogram frames are processed by a linear filter, which models the physical transfer of sound from the outer to the inner ear. The results of this operation are then treated by models for time and frequency masking. Masking is a psychoacoustic phenomenon in which the perception of a sound stimulus closely located to another in time or in frequency is changed due to their proximity. In PAQM, time masking is modeled by an exponential moving average over frames, with a different decay factor for each frequency bin. Frequency masking is modeled with two slopes in the bark scale that represents the mask spread across neighboring bins. The reader is invited to refer to~\cite{Beerends1992} for more details on both masking models. 

The resulting representations (dubbed excitation patterns in~\cite{Beerends1992}) are then compressed to account for the fact that perceived loudness is a nonlinear function of signal intensity, using a polynomial function with an exponent fitted in~\cite{Beerends1992} for the best correlation with an ISO/MPEG set of subjective audio quality evaluations.

After obtaining the internal representations of the reference and test signals, the test signal is separately scaled in the pitch ranges of $0$ to $2$ barks, $2$ to $22$ barks, and $22$ to $25$ barks by the ratio between the total energies of the reference and test signal within that pitch range. This ensures that the total loudness of reference and test are comparable.

The scaled test signal is then subtracted from the reference to obtain the raw PAQM score. Using a carefully fitted sigmoid function~\cite{Moreno21}, raw PAQM scores can be mapped to objective mean opinion scores (MOS) which are highly correlated to mean opinion scores obtained through subjective testing. Since PAQM is calculated using only differentiable operations, the correlation of PAQM objective MOS with subjective MOS suggests that it could be used as a loss function for NNs. 

With this in mind, the authors of this paper created \textit{torchpaqm}\footnote{https://github.com/bvm810/torchpaqm/tree/main}, a \textit{pytorch} based implementation of PAQM (as described in~\cite{Beerends1992}) that uses only \textit{pytorch} vector operations. Conversion from hertz to barks by filter bank was implemented as a matrix multiplication; outer to inner ear transfer was performed for all frames using broadcasting; and frequency masks were calculated with broadcasting for all bins and stacked in a single tensor before using the composite stimulus formula of~\cite{Beerends1992} to obtain the combined excitation pattern. The library also includes a \textit{pytorch} dataset class and a pre-made collator function that, when used together, allow the user to conveniently calculate PAQM scores for a batch of signals. 

The advantages of this implementation are two-fold. First, it allows using GPU acceleration to calculate PAQM scores for batches of reference and test signals, which provides significant speed gains and allows scaling up audio quality evaluation. Second, and most importantly, it enables the use of \textit{pytorch}'s automatic differentiation engine to obtain the derivatives of the PAQM scores with respect to a set of parameters. This in turn enables incorporating PAQM scores as a perceptual loss for a NN, which can have a particularly positive impact on the performance of super-resolution models, such as AERO.

\section{AEROMAMBA}
\begin{figure}[!ht]
    \centering
    \includegraphics[width=65mm]{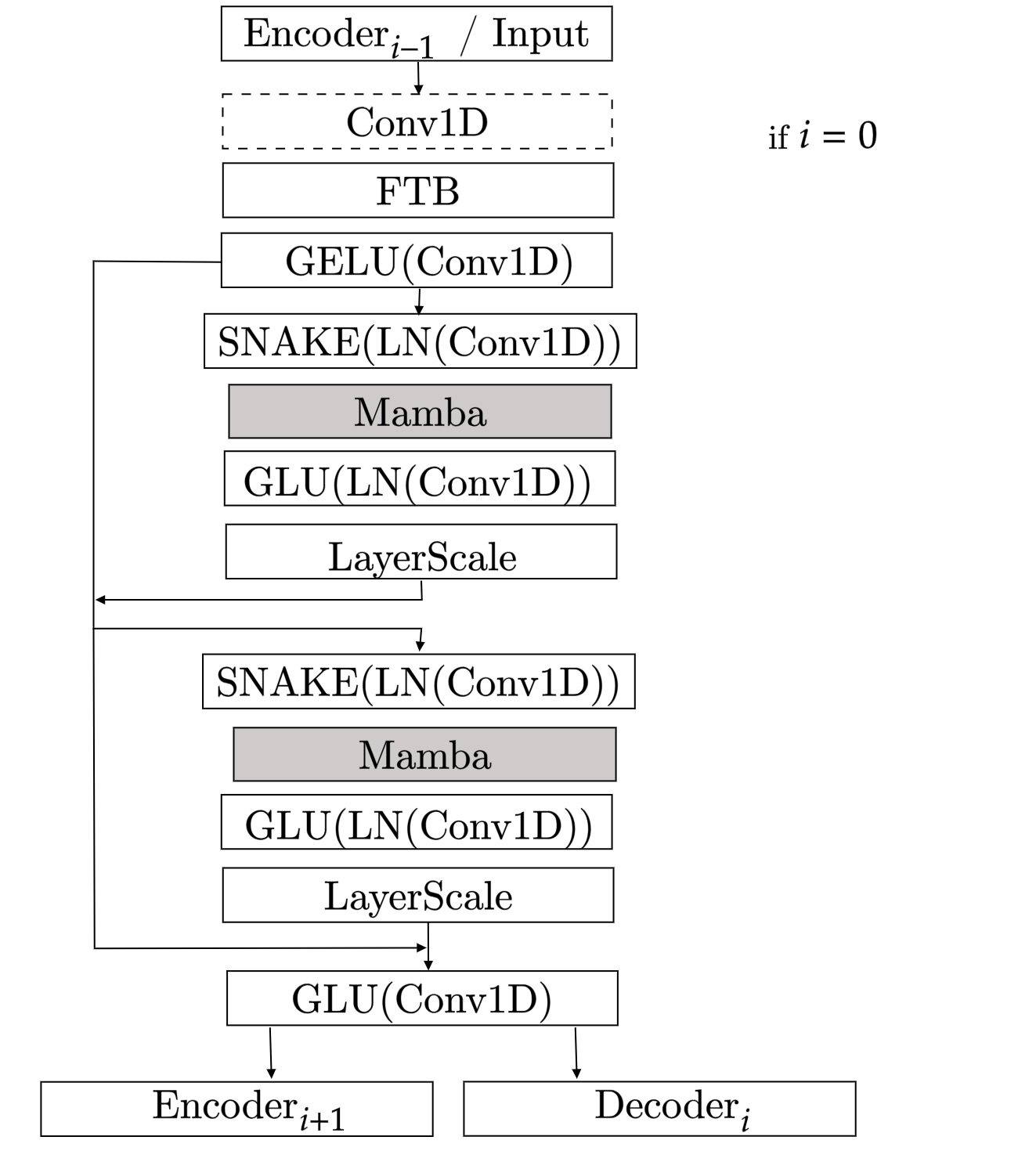}
    \caption{AEROMamba architecture.}
\label{fig:AEROMamba}
\end{figure}

Despite the state-of-the-art performance of AERO, both its output quality and computational efficiency can be improved. In this direction, the previous NN model \textbf{AEROMamba}~\cite{Abreu2024lamir} was designed to take advantage of Mamba selectiveness as a replacement for the attention mechanism. In AEROMamba, local attention and BiLSTM layers are replaced by Mamba layers and included in all encoder depths. The modified method performs the same tasks using reduced computational resources, as the Mamba architecture exploits the GPU memory hierarchy for optimized computations. The resulting architecture is shown in Figure~\ref{fig:AEROMamba}, with the added layer in gray.

This paper proposes \textbf{$\textrm{AEROMamba}_{\textrm{P}}$}, an improved version of AEROMamba that includes a perceptual component in its generator loss. In AERO and AEROMamba, the spectral reconstruction loss $L_\mathrm{rec}$ identifies energy distortions without evaluating whether these distortions are perceptually significant. Although the adversarial and feature map losses $L_\mathrm{adv}$ and $L_\mathrm{fmap}$ might act as perceptual proxies in some sense for these models, this behavior is not guaranteed. This motivated the idea of creating a model that explicitly incorporated a perceptual loss component. 

Specifically, a PAQM loss term $L_{\mathrm{PAQM}}$, standing for the MOS obtained from the application of PAQM to a pair of degraded and original signals, is incorporated into \eqref{eq:aero_g_loss}. The \textit{torchpaqm} package was used to obtain $L_{\mathrm{PAQM}}$. The modified loss $\tilde{L}_\mathcal{G}$ becomes 
\begin{equation}\label{eq:AEROMambapaqm_g_loss}
   \tilde{L}_\mathcal{G} = L_\text{adv} + L_\text{rec} + \lambda L_{\text{fmap}} - \gamma L_{\mathrm{PAQM}},
\end{equation}
where $\gamma$ is a scaling factor. Clearly, for $\gamma=0$ it aligns with the AEROMamba setting. Since $L_{\mathrm{PAQM}}$ produces values in the range $[1,5]$, and the goal is to minimize $\tilde{L}_\mathcal{G}$, the optimization procedure drives the model parameters to maximize $L_{\mathrm{PAQM}}$ and, as a consequence, increases the perceptual quality of the generated samples.

A particular variation of $\textrm{AEROMamba}_{\textrm{P}}$, called \textbf{$\textrm{AEROMamba}_{\textrm{P} \bar{\textrm{S}}}$}, is obtained when the reconstruction loss is effectively replaced by the PAQM loss, that is, a non-zero $\gamma$ is applied and the $L_\mathrm{rec}$ loss is completely removed, yielding 
\begin{equation}\label{eq:AEROMambapaqm_plusplus_g_loss}
   \tilde{L}_\mathcal{G} = L_\text{adv} + \lambda L_{\text{fmap}} - \gamma L_{\mathrm{PAQM}}.
\end{equation}
he motivation for using $\textrm{AEROMamba}_{\textrm{P} \bar{\textrm{S}}}$ to enhance encoded audio lies in two considerations: i) PAQM was originally conceived to evaluate lossy encoded audio; ii) a spectral reconstruction operation (based on mapping to an auditory model) is already implicit in $L_{\mathrm{PAQM}}$. It should be noted that, in addition to $L_{\mathrm{PAQM}}$, $\textrm{AEROMamba}_{\textrm{P} \bar{\textrm{S}}}$ also preserves the feature map and adversarial losses, which operate in the time domain to avoid phase artifacts.

\section{DATASETS}
\label{sec: datasets}

The experiments in this work used the two datasets described below. 
\subsection{PianoEval}
The PianoEval data set, collected by one of the authors, consists of two parts. Part 1 is composed of the 24 Preludes for Piano, \textit{op}. 28, by Chopin performed by 33 pianists in 45 different recordings available on CD (Compact Disc), totaling approximately 24 h. Part 2 contains excerpts of Ligeti piano études, a Schumann sonata, and Barber sonata, played by three different pianists, and totaling approximately 3.5 h. Each file is stored in WAV format, stereo mode, and sampled at 44.1 kHz. Information about performers, record label, and year of recording is detailed in the accompanying webpage.


\subsection{MUSDB}
The MUSDB~\cite{musdb18-hq} dataset is composed of 150 full-length music tracks of different genres along with their isolated drums, bass, vocals, and other stems, and totaling approximately 10 h of duration. The training folder contains 100 songs (around 7 h), while the test folder contains 50 songs (around 3 h), all of them in WAV format, stereo mode, and sampled at 44.1 kHz. The number of tracks per genre is distributed as follows: 83 Pop/Rock, 24 Rock, 15 Singer/Songwriter, 14 Heavy Metal, 13 Pop, 10 Rap, 9 Electronic, 3 Country, 3 Jazz, and 2 Reggae tracks.

\section{EVALUATION PROCEDURES}
The following evaluation procedures were applied to the two experiments considered in this paper, namely super-resolution of bandlimited audio and enhancement of heavily compressed audio, except for the efficiency evaluation, which was performed exclusively in the former.

\subsection{Efficiency evaluation}
The use of GPU by both AERO and the proposed methods was evaluated using the `nvidia-smi' command, with all models in \textit{pytorch}'s evaluation mode, and using different audio tracks. For inference time, the execution time for processing 10-s segments was recorded and the average duration for both methods was calculated. Since the architectural components of AEROMamba and its variants remain consistent within the experiments, the efficiency measures were only evaluated for AERO and AEROMamba within the bandlimited audio experiment.

\subsection{Objective evaluation}
\label{sec: objective-eval}

For objective quality evaluation, the log-spectral distance (LSD)~\cite{Mandel22} was used in combination with two distinct perceptual audio quality measures: PAQM (described in Section~\ref{sec:PAQM}) and ViSQOLAudio (Virtual Speech Quality Objective Listener extended for general audio)~\cite{Chinen20visqol}\footnote{referred to as ViSQOL in the tables}. PAQM served as a validation measure to select the optimal models during training, whereas ViSQOLAudio was used to evaluate the quality of the processed signals with respect to references in the test sets on a scale from 1 to 5. The latter is a necessary
impartial measure of quality, since PAQM is integrated into the $\textrm{AEROMamba}_{\textrm{P}}$ loss. Moreover, the findings suggested that comparing the quality of two signals with LSD is effective only when the values diverge significantly. Thus, ViSQOLAudio funcions as a more reliable measure of perceptual audio quality --- notably, it was also used in the AERO publication.

\subsection{Subjective evaluation}
Due to limitations of the objective metrics, a subjective evaluation test was designed to assess the quality of the outputs of each model. Among the different testing methodologies available, MUSHRA~\cite{MUSHRA} would have been the closest to our needs. However, our study required the evaluation of enhancement systems, which is not directly addressed in the standard MUSHRA framework, mainly focused on coding systems. Therefore, a specifically tailored evaluation methodology was designed. As detailed below, the structure of the listening test remains similar to that of MUSHRA, measuring fidelity to a high-quality reference, but with two main modifications: i) to reduce participant fatigue, only one anchor was used and no hidden reference was included; ii) the scoring scale was carefully designed to reduce biases induced by categorical labels or a numbered grid~\cite{Bech07}.

For each experiment, the opinion of 20 subjects was collected regarding the overall similarity of four test signals --- one corresponding to each model plus one anchor bandlimited to 5.5 kHz or encoded at 32 kbps, depending on the experiment --- to a known reference. 

Participants were recruited from students and faculty members affiliated with signal processing laboratories, a considerable number of whom had previous experience with auditory assessments. Specifically, more than half had over 5 years of experience working with audio. Participants' ages ranged from 20 to 61 years.

Experiments were carried out in an acoustically isolated environment within the Signals, Multimedia, and Telecommunications Laboratory at UFRJ, keeping consistent hardware and software configurations. The evaluation platform was accessed via a web browser, and participants listened to the audio using Sennheiser HD 265 linear headphones.

The score was assigned on an unnumbered sliding bar, with the words ``exactly the same’' on the right end and ``increasing difference'' indicated by a left arrow. The selected position was internally mapped to an integer from 0 to 100. 

Subjective testing was based on 12 tracks from the PianoEval dataset and 12 tracks from the MUSDB test set, detailed in the accompanying webpage. 
Each signal was evaluated by 10 subjects, and the order of the test questions was randomized to mitigate bias.

\subsection{Statistical significance}
All ViSQOL and subjective average scores were evaluated in pairs with Mann-Whitney statistical significance tests (due to the non-gaussianity of score distributions) and, except when explicitly stated, the mean values were considered to reject the null hypothesis of no significant difference with $p\mathrm{-value} < 0.05$. The statistics for each evaluation are tabulated in the accompanying webpage.

\section{SUPER-RESOLUTION OF BANDLIMITED AUDIO}

\subsection{Models}
In this experiment, different models were considered for subjective and objective evaluation. In the objective tests, AERO, AEROMamba, $\textrm{AEROMamba}_{\textrm{P}}$ and a pretrained version of AudioSR were evaluated by comparing their outputs with those of a simple bandlimiting process. In the subjective tests, AudioSR was not considered.

The pretrained AudioSR model served solely as a supplementary baseline, due to its distinct characteristics compared to the other architectures, such as its flexibility in handling various initial sampling frequencies and its extensive training of 7000 h (encompassing the entire MUSDB dataset). It was included in the objective evaluation using ViSQOL only to show how the other methods compared to 1) a much larger model that, 2) in the context of the MUSDB evaluation, had been exposed to the test data.

For $\textrm{AEROMamba}_{\textrm{P}}$, the PAQM loss weighting factor $\gamma$ was fixed without performing an extensive hyperparameter search. Although alternative weightings of the loss components might yield improved performance, this investigation was left for future work. Preliminary experiments with $\textrm{AEROMamba}_{\textrm{P} \bar{\textrm{S}}}$ in the context of super-resolution of bandlimited audio did not indicate perceptual advantages over $\textrm{AEROMamba}_{\textrm{P}}$; therefore, this variant was not included in the present analysis.

\subsection{Data}
This experiment used versions of AERO, AEROMamba and $\textrm{AEROMamba}_{\textrm{P}}$ trained on the datasets described in Section~\ref{sec: datasets}. Evaluation was later performed considering the test splits of these datasets (e.g. AERO trained on MUSDB was evaluated with the test split of MUSDB).

For evaluation with MUSDB, the original split was used, as in~\cite{Mandel22}: 100 tracks for training, and 50 for testing. Although MUSDB provides the isolated stems for each track, only the mixtures were used in the experiments below.

For evaluation with PianoEval, two subdatasets were created, and models were trained on each of them. The first subset, \textbf{PianoEval-GQ} (GQ for general quality) comprises all tracks recorded after 1950, while the second subset \textbf{PianoEval-HQ} (HQ for high quality) comprises all tracks recorded after 1960. Recordings prior to 1950 were absent from both subsets due to their very high noise levels. 

PianoEval-HQ contains the `HQ' tracks of Part 1 and all tracks of Part 2 of PianoEval; PianoEval-GQ adds the `non-HQ' tracks of Part 2 to PianoEval-HQ. This division of PianoEval aimed to take into account the impact of the intrinsic noise characteristic of older recordings during model training. For each architecture, versions trained on PianoEval-GQ and PianoEval-HQ were created, and the variant with the best performance, according to the objective evaluation, was retained for further subjective testing. In the results for PianoEval, models with the `HQ' suffix use PianoEval-HQ, while the others use PianoEval-GQ.

PianoEval-GQ and PianoEval-HQ were partitioned as follows: the same test set ($\approx 3.5$ h) containing all the pieces from Part 2 of PianoEval; the same validation set ($\approx 2$ h) containing `HQ' tracks from Part 1 of PianoEval; and training sets with the remaining 22 h of PianoEval-GQ or 20.5 h of PianoEval-HQ, respectively. 

\subsection{Training procedure}
As the baseline, AERO was trained using a window size of $W=512$ samples and a hop length of $H= 256$ samples, due to the computational requirements associated with training at hop sizes of 64 or 128 samples. This was not deemed a problem as, according to the original paper, AERO consistently outperforms competing methods when evaluated using the MUSDB dataset, even for the larger $H=256$. To ensure consistency across comparisons, AEROMamba and $\textrm{AEROMamba}_{\textrm{P}}$ were also trained using the same parameters employed for AERO, except for batch size. For $\textrm{AEROMamba}_{\textrm{P}}$, the PAQM loss scaling factor was set to $\gamma=1$. Non overlapping segments of 4 s (MUSDB) and 6 s (PianoEval, both) were used during training. Silent or excessively short segments were discarded.

All models were trained for $\approx800$ epochs. Model selection was determined by the convergence of the PAQM score on the validation set for PianoEval or on the test set for MUSDB\footnote{Although selecting the optimal model based on the test set is generally discouraged due to the risk of overfitting, the same methodology of the AERO paper was followed for consistency.}. The convergence criterion was defined as less than 0.01 variation in the PAQM score over 5 epochs. 
For AERO in MUSDB, a checkpoint after a pretraining of 696 epochs with batch size of 16~\cite{Mandel22} was used as a starting point; training was then resumed for 100 fine-tuning epochs, with a batch size of 4 due to GPU limitations. With the PianoEval subsets, AERO was consistently trained with a batch size of 4 throughout the 800 epochs. AEROMamba and $\textrm{AEROMamba}_{\textrm{P}}$ were trained along the 800 epochs with a batch size of 8 across all datasets, thanks to their reduced computational demands.

All relevant training parameters (except for $\gamma$, which only applies to $\textrm{AEROMamba}_{\textrm{P}}$) are summarized in Table~\ref{table:train_parameters_wav}; they were adopted in accordance with the AERO default settings, unless previously stated otherwise.
\begin{table}[htbp]
\centering
\renewcommand{\arraystretch}{1} 
\caption{Summary of key training parameters across datasets. $^*$ AERO fine-tuning; $^{**}$ Other models.}
\begin{tabular}{@{\hskip 2pt}l@{\hskip 6pt}c@{\hskip 6pt}c@{\hskip 2pt}} 
\toprule
\textbf{Parameter} & \textbf{MUSDB} & \textbf{PianoEval (GQ/HQ)} \\ \midrule
Window Size ($W$)     & 512 samples          & 512 samples          \\ 
Hop Length ($H$)      & 256 samples          & 256 samples          \\ 
Segment Length        & 4 s            & 6 s            \\ 
Stride Length         & 4 s            & 6 s            \\ 
Number of Epochs      & $\sim$800            & $\sim$800            \\ 
Batch Size            & $4 ^*$ / $8^{**}$ & $4 ^*$ / $8^{**}$ \\ 
Optimizer             & Adam~\cite{Kingma15} & Adam                 \\
Learning Rate         & $3 \times 10^{-4}$   & $3 \times 10^{-4}$   \\ 
\bottomrule
\end{tabular}
\label{table:train_parameters_wav}
\end{table}
\begin{table}[ht]
\centering
\caption{GPU usage (VRAM), inference time for a 10-s segment, and number of parameters for each model.}
\begin{tabular}{lccr}
\toprule
\textbf{Method} & \textbf{VRAM (MB)} & \textbf{Time (s)} & \textbf{Parameters} \\
\midrule
AERO & 17091 & 1.246 & 19,432,958 \\
AEROMamba & 3000 & 0.087 & 20,964,190 \\
\bottomrule
\end{tabular}
\label{tab:performance_comparison}
\end{table}

\subsection{Results}
\subsubsection{Efficiency evaluation}
The computational requirements of AERO and AEROMamba are compared in Table~\ref{tab:performance_comparison}, which reports GPU usage and time required to process a 10-s signal segment along with the number of parameters of each model. $\textrm{AEROMamba}_{\textrm{P}}$ was not included, since it uses the same backbone architecture as AEROMamba. Testing was performed with a NVIDIA RTX 3090 GPU.


\subsubsection{PianoEval}
As mentioned in Section~\ref{sec: objective-eval}, ViSQOL and LSD were used to compare references to test signals processed by each model. Each signal consisted of a 10-s excerpt from the respective track within the test dataset. Table~\ref{tab:pianoeval_scores} shows the average ViSQOL and LSD scores along with the corresponding average subjective scores attained by each method. The ``Low-Resolution'' line displays quality metrics for the bandlimited signals without any enhancement. 


The selection of models to be considered in the subjective evaluation followed a systematic two-step procedure: i) For each model, a statistical significance test (SST) was conducted comparing its ViSQOL scores with those of the low-resolution signals. Models for which no statistically significant difference (SSD) was observed were discarded.
ii) If both the HQ and non-HQ variants of a model passed this step, a second SST was performed between them. If there was an SSD, the variant with the higher ViSQOL score was chosen; otherwise, the LSD score was used as a tiebreaker. This procedure ensured that only the best models were selected for the subjective evaluation.
This procedure ensured that only the best models were selected for the subjective evaluation.
\begin{table}[!ht]
\centering
\caption{Objective and subjective scores for low-resolution signals and various models evaluated on PianoEval. $^*$Did not reject the null hypothesis compared to low-resolution.}
\begin{tabular}{lccc}
\toprule
\textbf{System} & \textbf{ViSQOL $\uparrow$} & \textbf{LSD $\downarrow$} & \textbf{Score $\uparrow$} \\
\midrule
Low-Resolution      & 4.36 & 1.09 & 72.92 \\
AERO                & 4.38 & 0.99 & 76.89 \\
AERO-HQ             & 4.34 & 1.04 & - \\
AEROMamba           & \,\,4.43$^*$  & 0.98 & - \\
AEROMamba-HQ        & 4.38 & 1.00 & \textbf{84.41} \\
$\textrm{AEROMamba}_{\textrm{P}}$      & 4.42 & 0.98 & - \\
$\textrm{AEROMamba}_{\textrm{P}}$-HQ   & 4.41 & 0.90 & 78.76 \\
AudioSR             & 3.89 & - & - \\
\bottomrule
\end{tabular}
\label{tab:pianoeval_scores}
\end{table}


\subsubsection{MUSDB}
As for PianoEval, Table~\ref{tab:musdb_scores} shows the average objective metrics along with the average subjective scores achieved by each method. 


\subsection{Discussion}

\subsubsection{Efficiency aspects}

From Table~\ref{tab:performance_comparison}, it is possible to see that Mamba allowed to reduce AERO's inference GPU requirements almost 6x while at the same time speeding inference almost 15x. Also, from the subjective and objective results in Tables~\ref{tab:pianoeval_scores}~and~\ref{tab:musdb_scores}, it is clear that the performances of Mamba variations of AERO (and of their HQ variants, for PianoEval) were superior or equal to AERO's. These performance improvements are due to the efficiency provided by the Mamba layer and also to its capacity in sequence modeling tasks. Furthermore, as seen in Table~\ref{tab:performance_comparison}, AEROMamba is a larger architecture in number of parameters that uses fewer computational resources, thus theoretically being a more powerful model. As an additional informal comparison, the same experiment using AudioSR with its default 50 diffusion steps resulted in an average inference time of 5.6 s, reflecting the slower sampling process inherent in diffusion models.
\begin{table}[!ht]
\centering
\caption{Objective and subjective scores for low-resolution signals and various models evaluated on MUSDB.}
\begin{tabular}{lccc}
\toprule
\textbf{System} & \textbf{ViSQOL $\uparrow$} & \textbf{LSD $\downarrow$} & \textbf{Score $\uparrow$} \\
\midrule
Low-Resolution      & 1.82 & 3.98 & 38.22 \\
AERO                & 2.90 & 1.34 & 60.03 \\
AEROMamba           & 2.93 & 1.23 & 66.47 \\
$\textrm{AEROMamba}_{\textrm{P}}$      & 3.04 & 1.19 & \textbf{79.26} \\
AudioSR            & 3.01 & - & - \\
\bottomrule
\end{tabular}
\label{tab:musdb_scores}
\end{table}

\subsubsection{PianoEval}

Results in Table~\ref{tab:pianoeval_scores} show that, in the objective evaluation, the scores were approximately equivalent across all systems. This outcome is attributed to the characteristics of the test data: the selected piano pieces exhibit, on average, a quiet, sparse, and delicate spectral-temporal structure. 

Considering the subjective scores of AEROMamba-HQ, AEROMamba is more effective in producing enhanced samples than AERO for piano tracks, and the incorporation of a PAQM loss is counterproductive in this setting. The statistical tests show that the only model that rejects the null hypothesis is AEROMamba-HQ. The considerable differences observed among the subjective scores of different systems suggest that the objective measures employed may lack sufficient sensitivity to capture perceptual differences in signals of this nature.

Qualitatively, AEROMamba-HQ was regarded by listeners as more prone to causing artifacts, which is a consequence of its tendency to generate higher frequencies even if it leads to an intensification of noise content. For piano signals, this was not necessarily detrimental to the musical result, sometimes bringing a brighter tone to the instrument. This was likely because the piano did not overload the model with too wide spectral content. As a consequence, the perceived quality of AEROMamba-HQ was deemed higher than its PAQM counterpart, which performed similarly to AERO.

An interesting observation is how AudioSR ViSQOL scores are lower than Low Resolution ones. This is caused by a common characteristic of perceptual evaluators: they penalize more severely the addition of spectral content not found in the reference than the removal of content (as in the filtered signal). Indeed, the pretrained AudioSR is prone to adding excessive percussive energy to its outputs, which also sounds unnatural for piano.

\subsubsection{MUSDB}
As seen in Table~\ref{tab:musdb_scores}, $\textrm{AEROMamba}_{\textrm{P}}$ outperformed the other models in all metrics on MUSDB. This is particularly noticeable for the subjective scores, and showcases the success of PAQM as a perceptual loss function. Intuitively, the effect of the PAQM loss term can be compared to that of a regularizer, steering the model in a direction that maximizes the perceptual quality, while balancing the generation of spectral content. With the inclusion of the PAQM loss term, the proposed model was able to achieve average ViSQOL results higher than those of AudioSR, which had previously been exposed to MUSDB. It should be noted, however, that the Mann-Whitney statistic comparing $\textrm{AEROMamba}_{\textrm{P}}$ and AudioSR for MUSDB had p-value higher than 5\%, as stated in the auxiliary web-page.

It is also interesting to see that in Table~\ref{tab:musdb_scores} signals processed by AERO and AEROMamba also rated much better than the Low Resolution signals. This discrepancy in relation to what was seen in PianoEval, where all scores were roughly the same, is due to the nature of the data: MUSDB tracks tend to span the entire spectrum and exhibit relatively stable high loudness, causing low-pass filtering to severely affect their overall quality.

\begin{figure}[!ht]
\centering
    \includegraphics[width=0.4\textwidth]{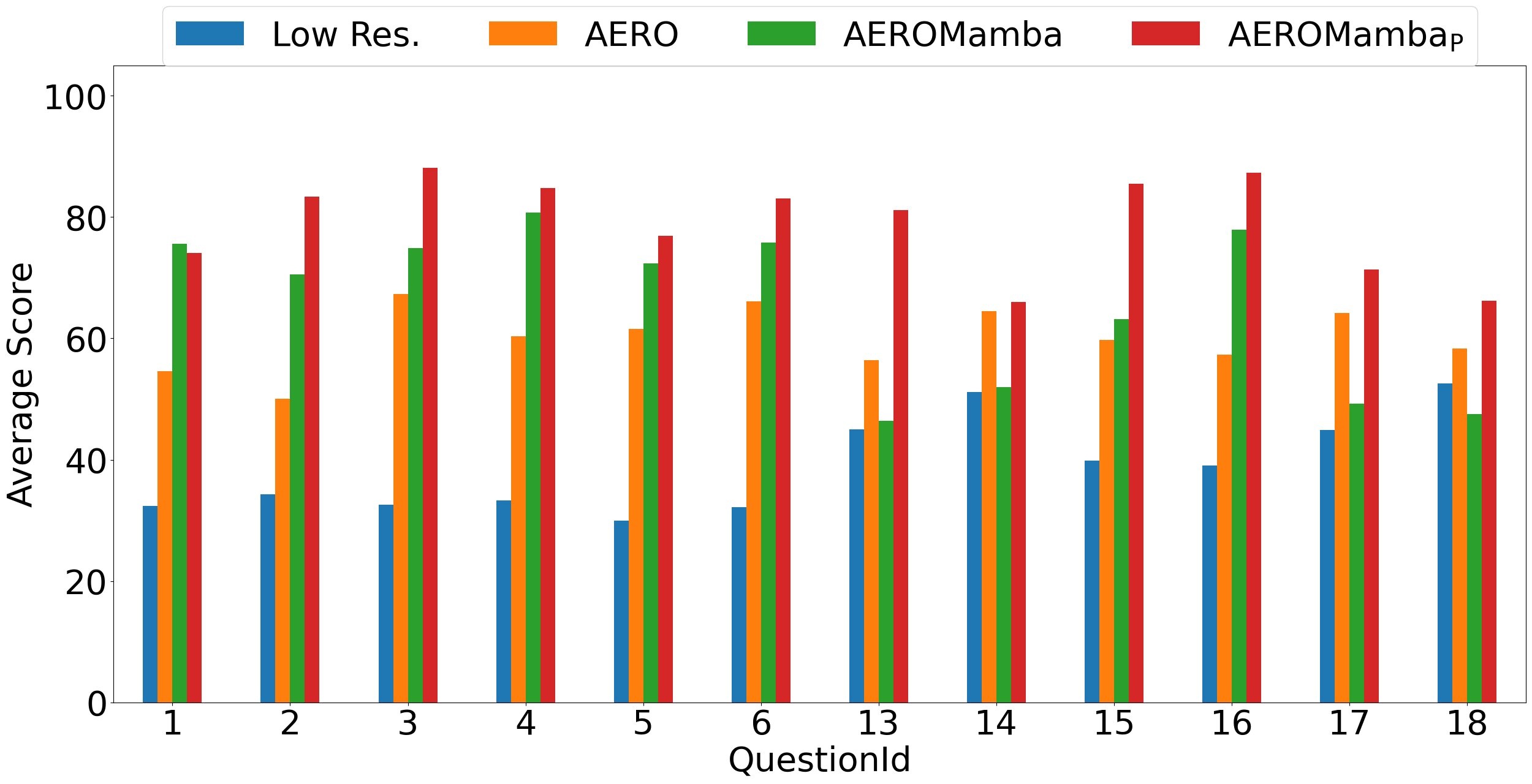}
    \caption{Subjective scores for low-resolution signals and various models per track of MUSDB.}
\label{fig:scores_bandlimited}
\end{figure}

For a detailed visualization, Figure~\ref{fig:scores_bandlimited} shows the scores for each track included in the subjective evaluation procedure, identified by their Id number on the testing interface. 

\begin{figure}[!th]
    \centering
    \includegraphics[width=6cm]{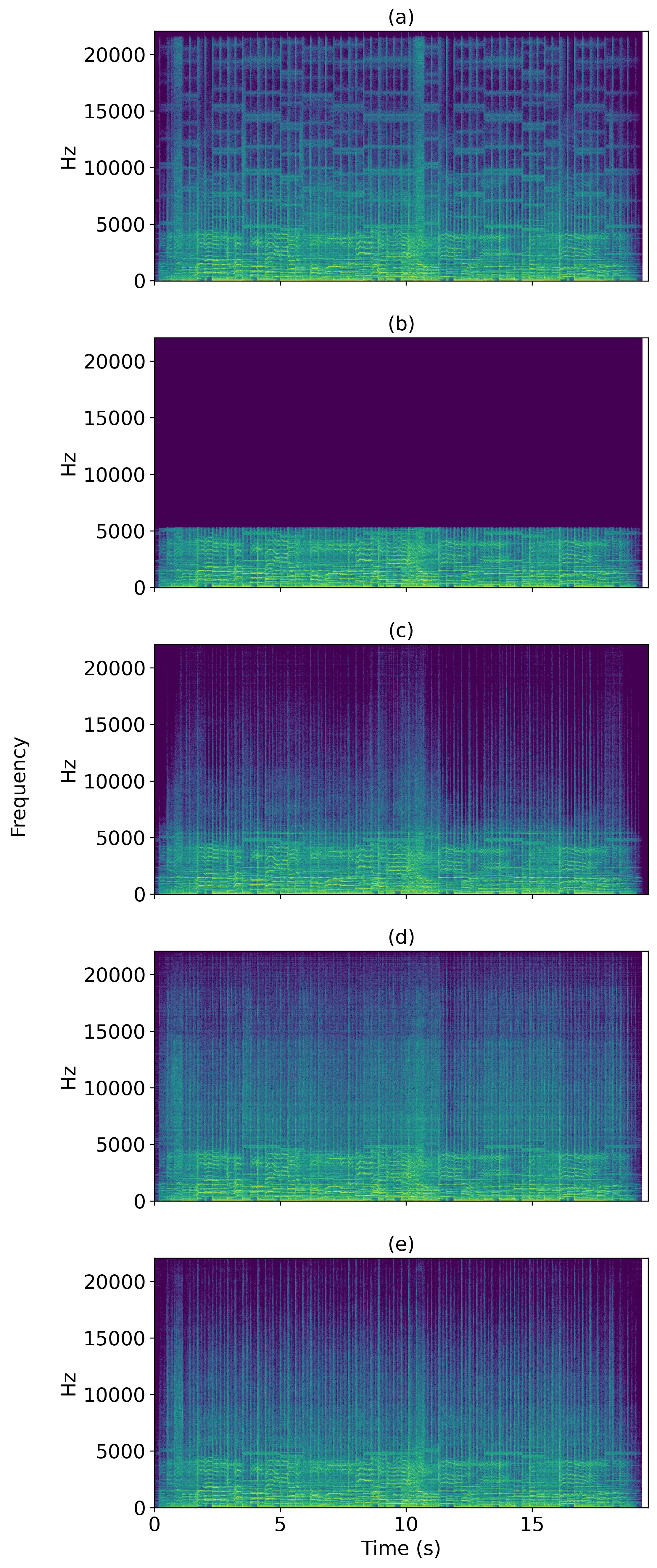}
    \caption[Distribution of subjective scores obtained from the evaluation of models on MUSDB tracks.]{Spectrograms for signal 15: (a) Reference, (b) Low-resolution, (c) AERO, (d) AEROMamba, and (e) $\textrm{AEROMamba}_{\textrm{P}}$.}
\label{fig:spectrogram_comparison_id15_vertical}
\end{figure}

For signal 15, on which $\textrm{AEROMamba}_{\textrm{P}}$ performed much better than the other methods, the spectra generated by the different processed signals are illustrated in Figure~\ref{fig:spectrogram_comparison_id15_vertical}. In this case, the reference signal occupies the full spectrum, including vocals delineated by singing patterns. The downsampling operation erases a considerable portion of frequency content, which is hard to recover for all models, especially in the vocals. However, $\textrm{AEROMamba}_{\textrm{P}}$ seems to achieve the most natural profile in the extended frequencies, with no clearly delimited regions in the extended range nor ghost spectral lines (which cause audible artifacts) as found in AERO or AEROMamba outputs.

\subsubsection{Concluding remarks}
In light of the obtained results,  it is possible to outline, in fairly general terms, some guidelines for choosing between the algorithms. (i) Taking the classical piano as a representative example of signals with predominantly tonal content, where the emphasis is on preserving timbre, the AEROMamba-HQ demonstrates greater freedom to generate more high-frequency content without creating artifacts —-- since the nature of the signal, with moderate transient characteristics, makes it less susceptible to this. (ii) In contrast, in popular music in general, the frequent occurrence of a rhythmic base, often based on percussion, demands greater precision from the system in transient events, which easily induce perceptible distortions. In this case, the additional "regularization" inherent in AEROMamba$_\mathrm{P}$ makes it more recommendable.

\section{ENHANCEMENT OF HEAVILY COMPRESSED AUDIO}
\subsection{Models}

AEROMamba, $\textrm{AEROMamba}_{\textrm{P}}$, and $\textrm{AEROMamba}_{\textrm{P} \bar{\textrm{S}}}$ are all evaluated in this experiment for the enhancement of heavily compressed audio. Their processed outputs are compared with a low-bitrate versions of reference signals, encoded using MP3.

\subsection{Data}
This experiment used the same datasets as the previous one, with the difference that only the GQ subset of PianoEval was used. This was motivated by the the fact that the degradations introduced by lossy coding can be considered to be more relevant than the presence of moderate noise in the case of heavily compressed audio.

Low-bitrate files were generated  by encoding the original files to MP3 format using FFmpeg~\cite{Tomar2006}). Since for a target bitrate of 32 kbps the encoder clips all spectral content of the audio above 5.5 kHz~\cite{LAME}, choosing this bitrate would allow us to investigate the applicability of the proposed solutions in the context of lossy compressed audio, without having to remove their current fixed cutoff frequency constraint. In addition to narrowing the spectral width, the compression operation also severely distorts the remaining part of the spectrum, causing artifacts such as spectral holes. The resulting decoded files were downsampled to 11.025 kHz.

\subsection{Training procedure}
Aside from the number of epochs, AEROMamba, $\textrm{AEROMamba}_{\textrm{P}}$, and $\textrm{AEROMamba}_{\textrm{P} \bar{\textrm{S}}}$ were trained with the parameters of Table~\ref{table:train_parameters_wav}, with $\gamma = 1$ for $\textrm{AEROMamba}_{\textrm{P}}$ and $\gamma=10$ for $\textrm{AEROMamba}_{\textrm{P} \bar{\textrm{S}}}$. The hypothesis behind this choice was that a greater weight on the perceptual cost could help the model achieve better performance, since PAQM was originally designed to evaluate the performance of lossy audio encoders. Models converged much faster in this experiment than in the previous, and so they were only trained for 200 epochs. The same criterion used previously for stopping training was adopted here.

\subsection{Results}
\subsubsection{PianoEval}

The objective metrics achieved for each method are shown together with the average subjective scores in Table~\ref{tab:visqol_lsd_comparison_mp3_piano}. 


\subsubsection{MUSDB}
As for the PianoEval case, the average objective metrics achieved for each method, together with the average subjective scores is shown in Table~\ref{tab:visqol_lsd_comparison_mp3_musdb}. 


\subsection{Discussion}

\subsubsection{PianoEval}
The subjective scores in Table~\ref{tab:visqol_lsd_comparison_mp3_piano} indicate that all models demonstrate nearly equivalent performance for the evaluated piano signals, exhibiting a subjective score improvement of approximately 14 points compared to the Low-Bitrate baseline. Notably, these scores closely resemble those achieved by AEROMamba-HQ in the preceding experiment. This suggests that the performance of the evaluated super-resolution tools on heavily compressed signals is comparable to that achieved for bandlimited audio. 
\begin{table}[!th]
\centering
\caption{Objective and subjective scores for the low-bitrate signals and various models models evaluated on PianoEval.}
\begin{tabular}{lccc}
\toprule
\textbf{System} & \textbf{ViSQOL $\uparrow$} & \textbf{LSD $\downarrow$} & \textbf{Score $\uparrow$} \\ 
\midrule
Low-Bitrate      & 4.35 & 2.33 & 69.5\\
AEROMamba           & 4.22 & 1.14 & 83.4\\
$\textrm{AEROMamba}_{\textrm{P}}$      & 4.24 & 1.12 & 84.1\\
$\textrm{AEROMamba}_{\textrm{P} \bar{\textrm{S}}}$    & 4.41 & 1.13 & \textbf{85.5}\\
\bottomrule
\end{tabular}
\label{tab:visqol_lsd_comparison_mp3_piano}
\end{table}
\begin{table}[!ht]
\centering
\caption{Objective and suibjective scores for low-bitrate signals and various models evaluated on MUSDB.}
\begin{tabular}{lccc}
\toprule
\textbf{System} & \textbf{ViSQOL $\uparrow$} & \textbf{LSD $\downarrow$} & \textbf{Score $\uparrow$} \\
\midrule
Low-Bitrate      & 1.80 & 2.02 & 50.7\\
AEROMamba           & 2.45 & 1.24 & 49.8\\
$\textrm{AEROMamba}_{\textrm{P}}$      & 2.99 & 1.27 & 49.7\\
$\textrm{AEROMamba}_{\textrm{P} \bar{\textrm{S}}}$    & 2.90 & 1.23 & \textbf{75.6}\\
\bottomrule
\end{tabular}
\label{tab:visqol_lsd_comparison_mp3_musdb}
\end{table}

The starting low-resolution and low-bitrate scores are also similar, which, in addition to the comparison above, indicates that for these tracks, heavy compression does not impose significantly worse perceptual degradation than linear filtering. Those distortions are more pronounced in segments where the signal has either high intensity or fast spectral variations, however, they represent small portions of the piano pieces. Therefore, $\textrm{AEROMamba}_{\textrm{P}}$ and $\textrm{AEROMamba}_{\textrm{P} \bar{\textrm{S}}}$ provide only incremental increases in subjective scores compared to AEROMamba. 

\subsubsection{MUSDB}
Looking at Table~\ref{tab:visqol_lsd_comparison_mp3_musdb}, it is possible to see that the starting low-bitrate scores for MUSDB were much lower than for PianoEval. As in the previous experiment, this can be explained by the fact that MUSDB signals tend to fill the spectrum much more than the piano-only signals of PianoEval; heavy MP3 compression introduces low-pass artifacts that are much more noticeable for MUSDB.

Another interesting observation is that AEROMamba and $\textrm{AEROMamba}_{\textrm{P}}$ show similar subjective scores to the low-bitrate signals, despite having higher ViSQOL scores. A possible explanation is that heavily compressed signals exhibit a of low-pass degradations and perceptually optimized non-linear distortions that might sound more or less natural sounding. The perceptual component of these degradations could have resulted in higher baseline subjective scores for the low-bitrate signals.

At the same time, the use of non-perceptual reconstruction loss terms in AEROMamba and $\textrm{AEROMamba}_{\textrm{P}}$ could have led these models to introduce audible artifacts in the processed signals while restoring high-frequency content. These artifacts were likely perceived during subjective evaluation, despite the extended bandwidth. The $\textrm{AEROMamba}_{\textrm{P} \bar{\textrm{S}}}$ variant, relying fully on a perceptually motivated frequency-domain loss, produces reconstructions that are more consistent with human auditory perception as per the subjective scores.

\begin{figure}[!ht]
\centering
\includegraphics[width=0.45\textwidth]{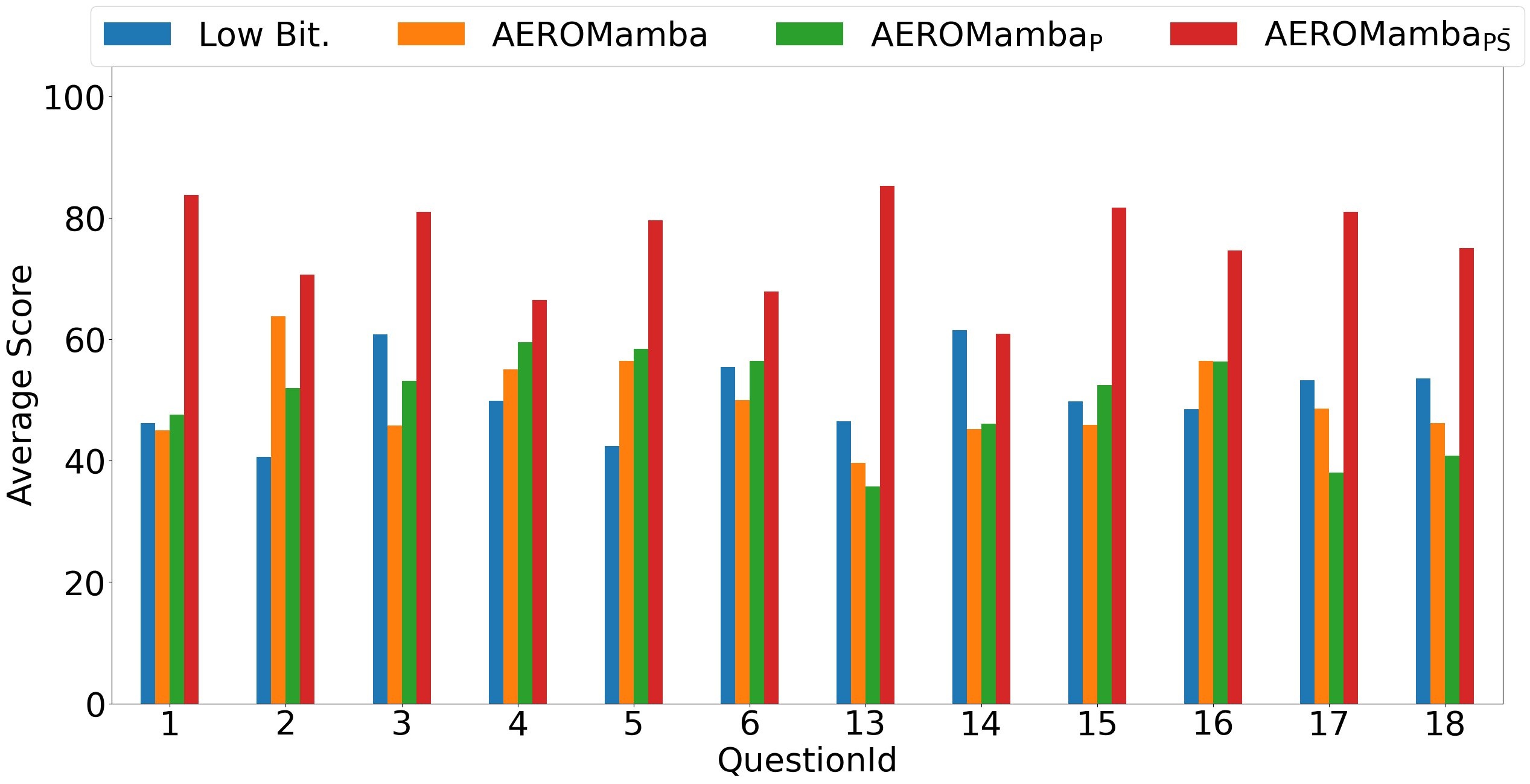}
\caption{Subjective scores for low-bitrate and various models per track of MUSDB.}
\label{fig:scores_mp3}
\end{figure}

Figure~\ref{fig:scores_mp3} shows the scores for each track included in the subjective evaluation procedure. 

\begin{figure}[!ht]
    \centering
    \includegraphics[width=6cm]{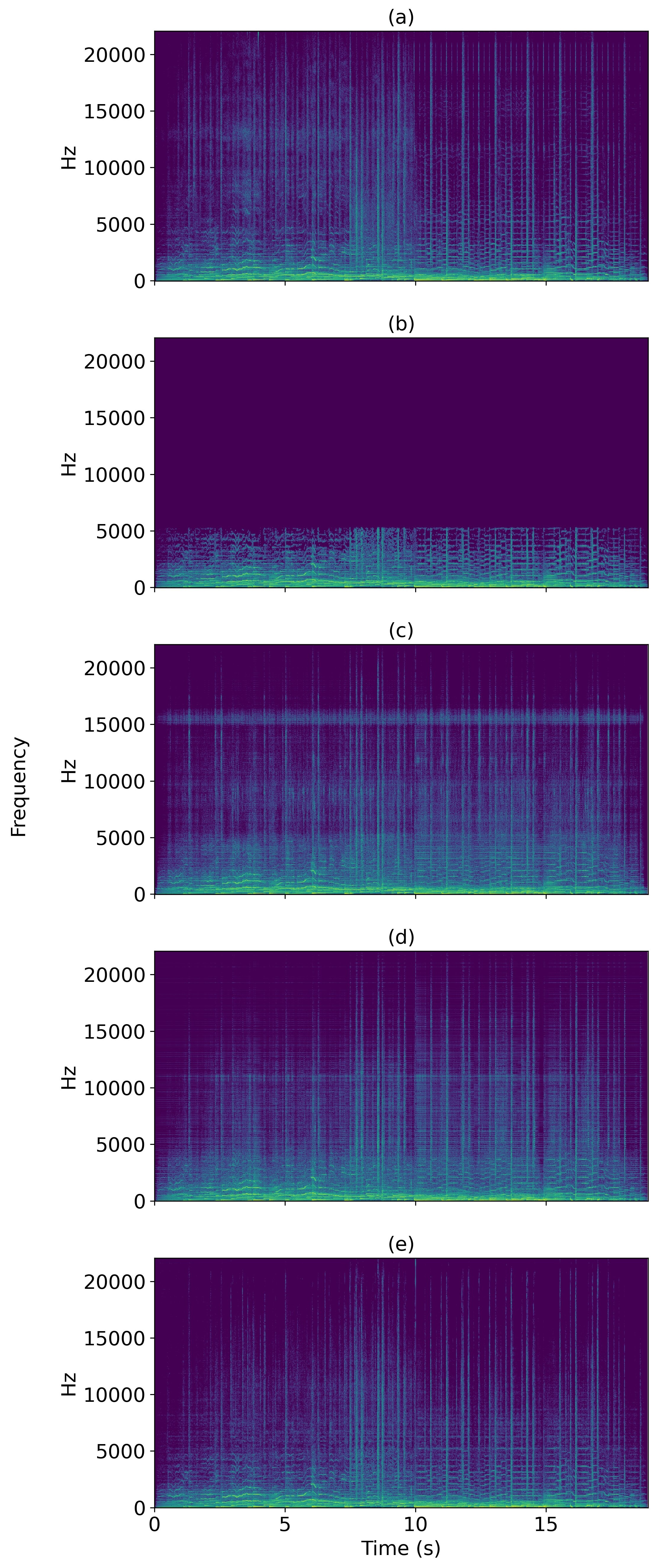}
    \caption{Spectrograms for signal 13: (a) Reference, (b) Low-bitrate, (c) AEROMamba, (d) $\textrm{AEROMamba}_{\textrm{P}}$, and (e) $\textrm{AEROMamba}_{\textrm{P} \bar{\textrm{S}}}$.}
\label{fig:spectrogram_comparison_id13_vertical}
\end{figure}

Track 13 was selected for visualization of spectra in Figure~\ref{fig:spectrogram_comparison_id13_vertical}, due to $\textrm{AEROMamba}_{\textrm{P} \bar{\textrm{S}}}$ performing much better than the other models.
The mentioned artifacts created by AEROMamba and $\textrm{AEROMamba}_{\textrm{P}}$ are visible in images (c) and (d), respectively. In contrast to the previous models, $\textrm{AEROMamba}_{\textrm{P} \bar{\textrm{S}}}$ is able both to reconstruct the low-frequency region and retrieve high-frequency content without creating noticeable anomalies, as shown in image (e). This can be attributed to the use of PAQM as the only frequency-domain loss, which is not only perceptually motivated, but also optimized to audio encoding scenarios, thus confirming the expectation of its efficacy in this scenario.

\subsubsection{Concluding remarks}
Experiments performed suggest that $\textrm{AEROMamba}_{\textrm{P}\overline{\mathrm{S}}}$ is the best choice for enhancing heavily compressed audio signals in general.


\section{CONCLUSION}
This paper proposes an efficient model for both bandlimited audio super-resolution and encoded audio enhancement using a perceptually motivated loss based on PAQM. It was shown, through objective and subjective metrics, that using this loss function can significantly improve the audio enhancement performance of the AEROMamba architecture in relation to other models, such as AERO and AudioSR, while at the same time keeping its computation efficiency. 

Audio quality improvements were consistent for both bandlimited and heavily compressed audio, especially in popular music, which is well represented by MUSDB. In particular, the experiments indicate that the commonly used STFT reconstruction loss, while useful for extending bandlimited audio, can be entirely replaced by a PAQM loss when addressing heavily compressed audio. It is worth noting that informal tests of super-resolution of old piano recordings suggest that including noisy recordings in training enables models to handle degraded material, to the point of performing denoising in conjunction with bandwidth extension --- an interesting topic for future investigation.

Next steps should include addressing the model limitation in operating only with a fixed oversampling factor, and performing a more thorough performance comparison with AudioSR and Apollo systems. Furthermore, since the STFT loss is used in many audio applications, a perceptually inspired loss can be effectively explored in scenarios other than bandwidth extension and inversion of lossy coding, e.g. audio restoration or music source separation. Finally, it is worth mentioning that PAQM is an effective and simple model of perceptual quality that is included in the larger standardized tool known as Perceptual Evaluation of Audio Quality (PEAQ)~\cite{Thiede00}. In future work, the latter should be implemented in \textit{pytorch} and evaluated as a more comprehensive perceptual cost for NN-based audio applications.

This work is accompanied by a code repository with training / inference scripts and model checkpoints, along with an online webpage (\url{aeromamba-paqm.github.io}) containing additional experimental material. The webpage provides playable audio examples, statistical test results, and score-distribution visualizations, supporting the transparency and reproducibility of the presented research.

\section{ACKNOWLEDGMENT}
This study was financed in part by the Coordenação de Aperfeiçoamento de Pessoal de Nível Superior – Brasil (CAPES) – Finance Code 001, the National Council for Scientific and Technological Development (CNPq) - Process \# 306395/2025-8, and the Carlos Chagas Filho Foundation for Research Support in the State of Rio de Janeiro (FAPERJ) - Grant \# E-26/204.092/2022).

\bibliography{jaes.bib}
\bibliographystyle{jaes.bst}

\end{document}